\documentclass[pdflatex,sn-apa]{sn-jnl}% APA Reference Style
\usepackage{makecell}
\usepackage{ragged2e}
\usepackage{hyperref}
\usepackage{graphicx}%
\usepackage{multirow}%
\usepackage{amsmath,amssymb,amsfonts}%
\usepackage{amsthm}%
\usepackage{mathrsfs}%
\usepackage[title]{appendix}%
\usepackage{xcolor}%
\usepackage{textcomp}%
\usepackage{manyfoot}%
\usepackage{booktabs}%
\usepackage{algorithm}%
\usepackage{algorithmicx}%
\usepackage{algpseudocode}%
\usepackage{listings}%
\theoremstyle{thmstyleone}%
\theoremstyle{thmstyletwo}%
\theoremstyle{thmstylethree}%
\author[1]{\fnm{Kun} \sur{Liu}}\email{2023210967@email.cufe.edu.cn}

\author*[2]{\fnm{Yan} \sur{Zhang}}\email{15420210156591@stu.xmu.edu.cn}
\author*[1]{\fnm{Rui} \sur{Pan}}\email{panrui\_cufe@126.com}
\author[2]{\fnm{Tianchen} \sur{Gao}}\email{gaotianchen@stu.xmu.edu.cn}
\author[3]{\fnm{Hansheng} \sur{Wang}}\email{hansheng@pku.edu.cn}

\affil[1]{\orgdiv{School of Statistics and Mathematics}, \orgname{Central University of Finance and Economics}, \orgaddress{ \state{Beijing}, \country{China}}}

\affil[2]{\orgdiv{School of Economics}, \orgname{Xiamen University}, \orgaddress{\city{Xiamen}, \state{Fujian}, \country{China}}}

\affil[3]{\orgdiv{Guanghua School of Management}, \orgname{Peking University}, \orgaddress{\state{Beijing}, \country{China}}}

\begin{document}

\title[Academic Literature Recommendation in Large-scale Citation Networks Enhanced by Large Language Models]{Academic Literature Recommendation in Large-scale Citation Networks Enhanced by Large Language Models}

\abstract{
Literature recommendation is essential for researchers to find relevant articles in an ever-growing academic field. However, traditional methods often struggle due to data limitations and methodological challenges. In this work, we construct a large citation network and propose a hybrid recommendation framework for scientific article recommendation. Specifically, the citation network contains 190,381 articles from 70 journals, covering statistics, econometrics, and computer science, spanning from 1981 to 2022. The recommendation mechanism integrates network-based citation patterns with content-based semantic similarities.
To enhance content-based recommendations, we employ text-embedding-3-small model of OpenAI to generate an embedding vector for the abstract of each article. The model has two key advantages: computational efficiency and embedding stability during incremental updates, which is crucial for handling dynamic academic databases.
Additionally, the recommendation mechanism is designed to allow users to adjust weights according to their preferences, providing flexibility and personalization.
Extensive experiments have been conducted to verify the effectiveness of our approach. In summary, our work not only provides a complete data system for building and analyzing citation networks, but also introduces a practical recommendation method that helps researchers navigate the growing volume of academic literature, making it easier to find the most relevant and influential articles in the era of information overload.
}


\maketitle

\section{Introduction}\label{sec1}

The citation network is a typical type of academic network, where nodes represent published articles, and edges represent citation relationships. Citation networks are widely studied in various disciplines, including physics \citep{newman2008physics},  computer science \citep{zhu2015dynamic}, statistics \citep{ji2016coauthorship}, bibliometrics \citep{zhao2022utilizing}, and many others. This type of network is extremely valuable for assessing the value of academic publications \citep{zhou2022influence} and identifying influential articles \citep{zhao2022utilizing}. By exploring the structure of citation networks, one can identify research topics \citep{ji2016coauthorship}, trace their evolution, and predict future research trends \citep{gao2024community}. In addition to these advantages, citation networks have a special role in academic recommendation. References in an article are more or less related to the content of that article. When reading an article, encountering a particularly interesting citation often prompts a search for the original source of that reference. Consequently, citation relationships between academic articles create an effective mechanism for locating the desired articles. As a result, the citation network established through these relationships serves as a valuable tool to make academic recommendations \citep{pornprasit2022enhancing}. Through this network, we can not only easily identify which articles an article has cited, but also discover which articles have cited that article.  

The construction of a citation network with a strong role in academic recommendation requires the collection of a large amount of citation information between articles. Some open resource citation networks can be found in the existing literature. For example, \cite{ji2022co} collected 83,331 articles published in 36 representative journals in statistics, probability, machine learning, and related fields from 1975-2015, along with their citation relationships. In addition, \cite{Tang:08KDD} is continuously collecting citation data from DBLP (\href{https://dblp.org}{https://dblp.org}), owning 5,259,858 articles and 36,630,661 citation relationships until 31 January 2023. The articles in this dataset come from major computer science journals and proceedings. Moreover, INSPIRE (\href{https://inspirehep.net}{https://inspirehep.net}), serves as a one-stop information platform for the high energy physics community, comprising eight interlinked databases on literature, conferences, institutions, journals, researchers, experiments, jobs, and data. INSPIRE contains a total of 1,305,067 articles and 34,893,469 citation relationships. 
In this work, we expand the dataset collected in the previous work of our team \citep{gao2023large} to 70 journals and construct a larger citation network. The 70 journals cover research in fields including statistics, econometrics, and computer science, with a particular emphasis on data mining. The list of the journals is shown in Table \ref{Tab:journals}. Specifically, our citation network consists of 190,381 articles and 1,087,277 citation relationships spanning from 1981 to 2022.  Our citation network data and article information data are publicly available at \href{https://github.com/Gaotianchen97/Academic-Literature-Recommendation-in-Large-scale-Citation-Networks}{https://github.com/Gaotianchen97/Academic-Literature-Recommendation-in-Large-scale-Citation-Networks}. Unlike the publicly available citation network mentioned above, we also obtain the abstract of each article and further embed the abstract texts to vectors via a large language model (i.e., the text-embedding-3-small model interface provided by OpenAI GPT). The model is highly stable, enabling efficient embedding of new article abstracts without requiring re-embedding of existing ones, ensuring scalability and reduced computational cost as the dataset expands. The use of large language models in literature recommendation systems is becoming a prominent trend, as demonstrated by tools like RefAI \citep{li2024refai} and ScholarAI (\href{https://scholarai.io/}{https://scholarai.io/}). This trend highlights the growing recognition of the use of advanced text embeddings to enhance the relevance and quality of academic recommendations.

With so much citation data available, researchers find it challenging to navigate the large and constantly growing amount of academic literature to find the most relevant and important articles in their field \citep{kanwal2024research}. Therefore, it is essential to find a method for searching and recommending articles effectively. Traditional methods of literature search, such as keyword-based search engines, often return large numbers of irrelevant results, making it time-consuming and difficult for researchers to find the articles that are most relevant to their work. To address this challenge, there is a need for more sophisticated and intelligent approaches to literature recommendation that can help researchers identify the most important and influential articles in their field. Various literature recommendation methods have been proposed in recent years \citep{sterling2021combining,pornprasit2022enhancing,li2024refai}. They can be roughly classified into four categories, i.e., user-based, content-based, network-based, and hybrid-based methods. Among them, user-based methods utilize the web usage log of previous searchers to recommend relevant articles \citep{hwang2003prototype}. Content-based methods calculate various measurements of textual similarity \citep{chen2019innovative,sterling2021combining}. Network-based methods construct various networks and analyze different similarity measurements \citep{haruna2017collaborative,kanwal2024research}. Hybrid-based methods are weighted combinations of two or three of the above methods \citep{liu2021applying}. We will review and compare the above methods in detail in Section \ref{sec2}.

In this work, we propose a new hybrid-based method of literature recommendation that combines citation patterns with textual similarities in abstracts and titles. The core idea is that citation patterns can reveal the relevance and significance of an article within its field, while the semantic similarity between abstracts and titles provides additional insights into the content of articles. In particular, we use large language models to generate text embeddings, which allow us to capture rich, context-sensitive relationships between articles. The rest of the work is organized as follows. Section \ref{sec2} reviews the related work. Section \ref{sec3} introduces our dataset in detail. The proposed hybrid-based recommendation method is presented in Section \ref{sec4}. The experiment with our real dataset is conducted to evaluate our method in Section \ref{sec5}. Finally, we conclude the work with a brief discussion in Section \ref{sec6}.

\section{Related Work}\label{sec2}

This section presents a review of four article recommendation methods, i.e., user-based, content-based, network-based, and hybrid-based methods. Among these methods, user-based methods were the first to emerge and have been used regularly in the domain of academic article recommendation \citep{thorat2015survey}. The other three methods have been widely used in recent studies \citep{haruna2018citation,sterling2021combining,liu2021applying,li2024refai}.

\subsection{User-based Recommendation}

The initial methodology used to recommend scientific articles is based on databases and user logs \citep{bai2019scientific}. Recommendation results are updated and sorted by collecting feedback from user clicks and browsing behavior. This method is also known as collaborative filtering \citep{goldberg1992using,resnick1994grouplens}, which implies that recommendations are made and improved through user participation \citep{beel2016paper}. 
As an illustration, \cite{hwang2003prototype} utilize the web usage logs and propose a recommendation framework that consists of three consecutive steps: preparing the web usage log data, discovering association of articles and recommending articles. Publications are ranked according to the preferences of active users. 
Similarly, \cite{lai2013cross} propose a cross-language literature recommendation system that utilizes weblog data and incorporates keyword translations to mitigate linguistic discrepancies. 
However, collaborative filtering cannot produce accurate recommendations when there are not enough initial user ratings \citep{thorat2015survey}. Even when we obtain a large volume of user logs, it is computationally intensive to process them to update recommended articles \citep{shi2014collaborative}. In addition, there are concerns about privacy and channels with regard to the acquisition of user logs \citep{calandrino2011you}.

\subsection{Content-based Recommendation}

Since the main information in academic articles is concentrated in the text, textual content-based recommendation methods have become the focus of many researchers \citep{huang2014refseer,bhagavatula2018content,sterling2021combining}.
A classic work in this field is RefSeer, a Latent Dirichlet Allocation (LDA)-based article recommendation system proposed by \cite{huang2014refseer}. RefSeer can infer the top $5$ topics of the content of the query article and recommend articles for each topic using the Cite-PLSA-LDA model. The evaluation results show that this system can recommend citations with good precision and recall.
Another method of utilizing information from textual content is text vectorization. For example, \cite{bhagavatula2018content} propose a recommendation method that uses a neural model to encode the textual content of each document, embedding all available documents in a vector space. The nearest neighbors of the given query article are considered as candidates for recommendation.
Similarly, \cite{chen2019innovative} utilize the frequency-inverse document frequency (TF-IDF) and WordNet techniques for text vectorization. 

However, the main limitation of the content-based recommendation methods is that they assume that the full content of the recommended articles is freely accessible and available \citep{haruna2017collaborative}. In practice, there are problems and difficulties in accessing and using the full text of the articles due to copyright restrictions \citep{laakso2016journal}. An alternative solution is to use text that is more readily accessible and free of copyright issues, such as titles, abstracts, citation information, and others. These are publicly available in databases such as the Web of Science (WoS) and journal websites.
For instance, \cite{sterling2021combining} utilize the textual similarity of the citation information and abstracts to provide a set of highly relevant articles. The system will search for articles in databases that are closely related to the set of query articles. Two scientific databases are used, namely OpenCitations \citep{peroni2020opencitations} and the Crossref API \citep{lammey2015crossref}. It will initially query the OpenCitations database and if the article is not found therein, it will then consult the Crossref API. This recommendation system is used to recreate references for review articles to demonstrate its effectiveness.

\subsection{Network-based Recommendation}

Regarding network-based recommendation, the definition and calculation of node similarity are the main focus. Many researchers use the Pearson correlation to measure node similarity in the early stages \citep{mccain1990mapping}. However, the suitability of the Pearson correlation is questioned in \cite{ahlgren2003requirements,van2008appropriate} and demonstrated to be an ineffective measure of node similarity in co-citation relationships between authors. Alternatives based on network structure are then proposed, e.g., the number of co-citations, cosine similarity between nodes, KL scatter, and Bhattacharyya distance. Furthermore, \cite{haruna2018citation} construct an article citation matrix, using Jaccard similarity to recommend relevant articles. 
In addition, scholars continue to define new measurements to better calculate node similarity. To illustrate, \cite{lu2007node} present two variants of node similarity, i.e., separate local neighborhoods and joint local neighborhoods. The results show that there is a complementarity between content-based and network-based similarity, providing a basis for subsequent hybrid approaches. 
Scholars also focus on the combination of various similarities. \cite{nassiri2013normalized} consider a citation network and propose a normalized similarity measurement based on shared common citations and references. The results show that the combined similarities can prioritize citations to highly cited articles, reflecting the idea of assigning weights to different articles for the first time. Recently, \cite{kanwal2024research} propose a recommendation method that combines both the measurements calculated from article citation network and author collaboration network. Specifically, they identify key authors for articles to determine a quality measure to recommend the top $n$ high-quality articles.

In addition to defining node similarity, the set of candidate articles from which similarity is calculated is also worth examining. If a recommendation is made from the entire database, it is computationally intensive to calculate the similarity between the target article and other articles. Therefore, a suitable source range of recommended articles is meaningful. 
Based on this idea, \cite{haruna2017collaborative} propose to identify the potential associations between articles based on the citation relationship between articles. They argue that the similarity between the target article and candidate articles needs to be investigated and weighed when their relationship is more closely defined. That is, they only calculate the similarity of a candidate $A$ if it cites any reference to the target, and there is another candidate that cites both the candidate $A$ and the target. This study provides the idea of calculating similarity in the partial structure in the network, which can reduce arithmetic and time consumption in similarity computation and recommendation.

\subsection{Hybrid-based Recommendation}

Hybrid methods can be simply regarded as a combination of the methods mentioned above. Some hybrid methods mix user-based recommendation methods with two other recommendation methods. Specifically, \cite{yang2013task} combine information retrieval, common citation analysis, and co-authorship analysis techniques with a citation network analysis technique to make recommendations. The results show that the hybrid method outperforms traditional user-based recommendation methods. Similarly, \cite{labille2015conceptual} recommend articles based on the similarity between articles and the conceptual profile of users, using both document text content and user information. 
However, limited by the shortcomings of user-based methods such as user log access, these hybrid methods are rarely used today.

In contrast, the findings of \cite{lu2007node} suggest a complementary between content-based and network-based similarity, providing a rationale for a hybrid method using these two methods. 
As an extension, \cite{boyack2010co} propose a hybrid method based on bibliographic coupling, showing advantages over pure network-based methods in all respects. Similarly, \cite{hamedani2016simcc} also consider both content and the application of a novel relevance and authority weighting strategy enables the method to effectively recommend articles that exhibit both content similarity and authority. \cite{hwang2017coauthorship} propose a hybrid method using LDA to integrate topic information into the links of the co-authorship network, achieving greater or comparable recommendation effectiveness compared to the pure content-based method. 
Recent research further takes into account the range of articles to the recommended articles in particular network structure. For example, \cite{liu2021applying} use a combination of TF-IDF and cosine similarity to calculate the similarity of citation content between co-citation structures in the citation network.

In addition to a hybrid of three traditional methods, the latest research is beginning to incorporate AI-assisted recommendations. \cite{li2024refai} propose the RefAI system, using PubMed for systematic literature search and GPT-4-turbo to assist recommendation. The evaluation results from domain experts show the role of large language modeling in helping the article recommendation system. Inspired by these important works, our recommendation system allows the user to choose one or both of content-based and network-based recommendation methods by assigning different weights. We also use AI tools to assist in text vectorization. This design aims to overcome the disadvantages of using only one recommendation technique and to embrace the advantages of hybrid methods.

\section{Datasets}\label{sec3}

The data used in this research are obtained from the WoS, which provides a database for global access to academic information. The WoS website (\href{https://www.webofscience.com}{https://www.webofscience.com}) allows researchers to search the literature in the direction of the natural sciences, social sciences, and many others. Specifically, we collect articles published in 70 specified journals between 1981 and 2022. The list of the journals is shown in Table \ref{Tab:journals}. The journals can be mainly divided into three categories, i.e., statistics, econometrics, and computer science. These three categories are the main directions for the development and integration of statistics, and they are also three fields of particular interest to statisticians.
Journals in the field of statistics include the four recognized top journals, i.e., the Annals of Statistics (AoS), Biometrika (Bka), Journal of the American Statistical Association (JASA) and Journal of the Royal Statistical Society Series B: Statistical Methdology (JRSS-B). Journals on econometrics include Econometrica and Journal of Econometrics (JoE). In addition, journals of computer science contain some high-quality journals in the IEEE Transaction Series and Journal of Machine Learning Research (JMLR).
We crawl a total of 190,381 articles published in the above mentioned journals. A distinctive ID is assigned to each article to facilitate its unique identification. For each article, we obtain its title, authors, publisher, publication year, abstract, keywords, citation count, number of references, and reference list. To present our dataset more clearly, we choose the article titled ``The adaptive lasso and its oracle properties'' to show its information, which is presented in Table \ref{table:data sample}.

\begin{table}[!htbp] 
\caption{One example from our dataset. Each article has a unique ID, along with its title, authors, publisher, publication year, abstract, keywords, citation count, number of references, and reference list.}
\label{table:data sample}
\begin{tabular}{ll} 
\toprule
Variables & Value\\ 
\midrule ID & paper\_90296\\
Title & The adaptive lasso and its oracle properties\\
Author & Zou, Hui\\
Publisher & Journal of the American Statistical Association\\
Publication year & 2006\\
Abstract & 
\makecell[l]{The lasso is a popular technique for simultaneous estimation and variable\\ selection. Lasso variable selection has been shown to be consistent under\\ certain conditions. In this work we derive a necessary condition for the l-\\asso variable selection to be consistent. Consequently, there exist certain\\ scenarios where the lasso is inconsistent for variable selection\dots}\\
Keywords & \makecell[l]{asymptotic normality; lasso; minimax; oracle inequality; oracle procedure; \\variable selection}\\
Citation count & 3,630\\
\makecell[l]{Number of \\references} & 27\\
Reference list & \makecell[l]{Regularization of wavelet approximations. Antoniadis, A., \& Fan, J., 2001, \\Journal of the American Statistical Association.\\ Heuristics of instability and stabilization in model selection. Breiman, L., \\1996, The Annals of Statistics.\\ Atomic decomposition by basis pursuit. Chen, S. S., Donoho, D. L., \& \\Saunders, M. A., 2001, SIAM Review.\\ Ideal spacial adaptation by wavelet shrinkage. Donoho, D. L., \& Johnstone,\\ I. M., 1994, Biometrika. \\ Uncertainty principles and ideal atomic decomposition Donoho, D. L., \& \\Huo, X., 2001, IEEE Transactions on Information Theory.\\ \dots } \\
\bottomrule 
\end{tabular}
\end{table}

\begin{figure}[!htbp]
\centering
\includegraphics[scale=0.27]{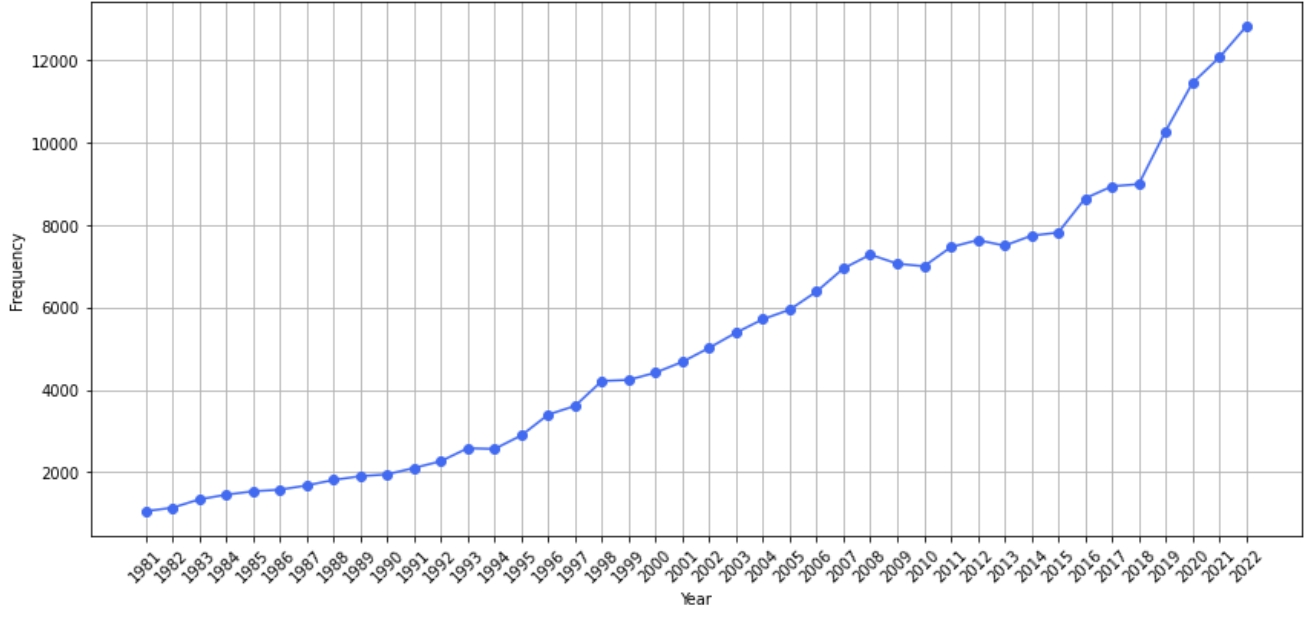}
\caption{Line graph of the change in the number of articles published from 1981 to 2022 for articles included in our dataset. The number of articles published annually demonstrates a flat growth pattern until 2018, followed by a more accelerated growth trajectory in subsequent years.}
\label{fig:publish_num}
\end{figure}

We present a line graph in Figure \ref{fig:publish_num} that illustrates the number of articles published annually from 1981 to 2022. A clear upward trend can be observed over the last 41 years. A similar phenomenon has also been observed in other research fields \citep{bornmann2015growth,collyer2018global,zhang2021will,savage2022more}. The increase in the number of academic articles can be attributed to several factors, including the increase in the number of academic researchers, improvements in research conditions and efficiency, policy incentives and assessment requirements, as well as interdisciplinary cross-fertilization and integration \citep{stephens2013methods,tonta2018does}. Due to the huge number of academic articles, scholars need to put a lot of effort into finding the articles they are interested in \citep{kanwal2024research}. Academic search engines such as Google Scholar and Baidu Scholar are widely used databases that span across academic disciplines. However, they sometimes suffer from the difficulty called academic search engine spam \citep{beel2010academic}. For example, Google Scholar indexes invisible text added to certain articles, causing these articles to appear in searches for unrelated keywords. The existence of spam in academic search engines may hinder scholars from effectively searching for relevant information. Therefore, it is necessary to design an effective article recommendation mechanism to recommend articles in specific fields that scholars are truly interested in. Predictably, this mechanism for recommending articles can improve the efficiency of article retrieval.

\begin{figure}[!ht]
\centering
\includegraphics[scale=0.35]{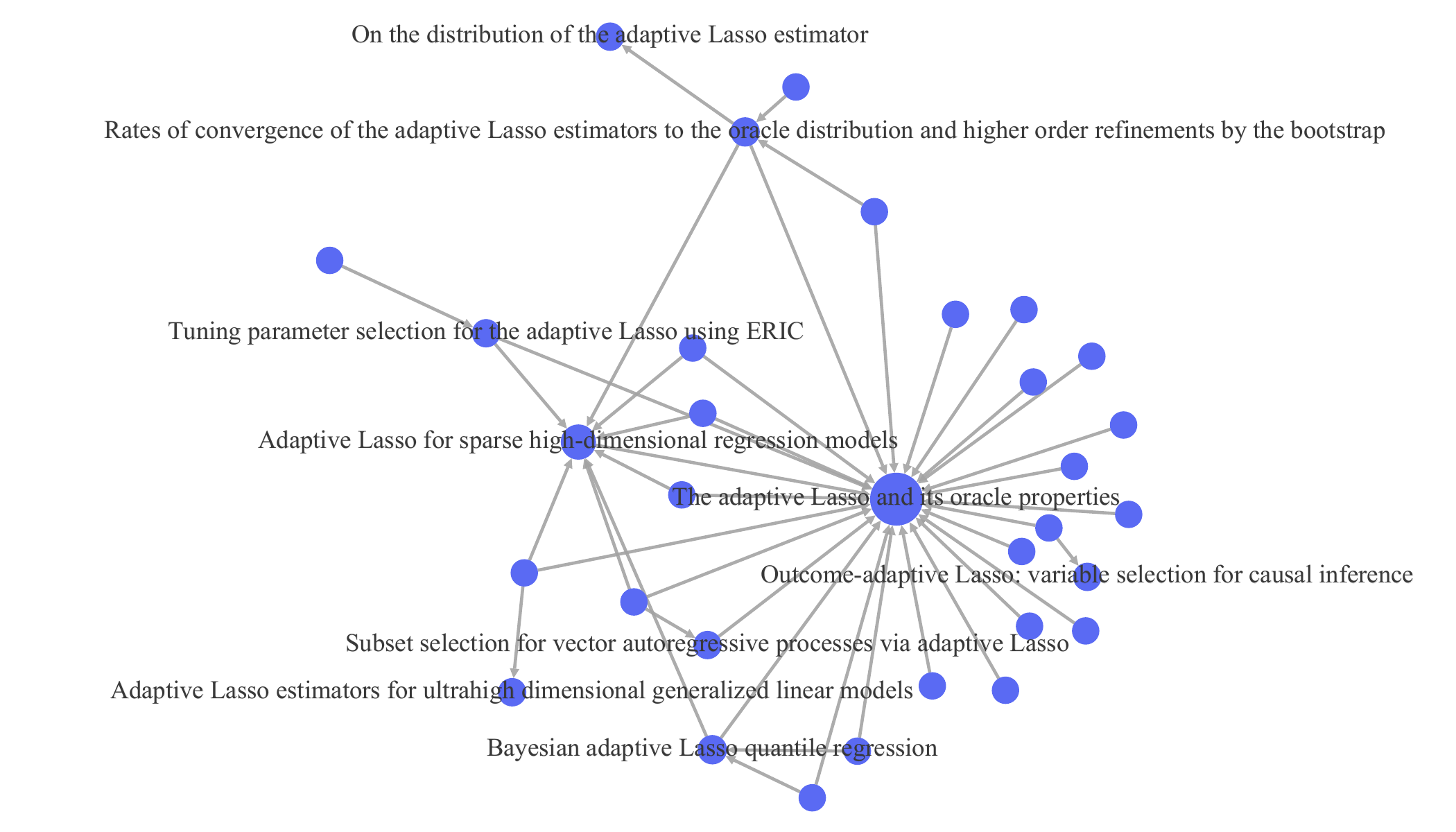}
\caption{Schematic diagram of the construction and components of a citation network. Each node represents an article, and arrows pointing from the start node to the arrival node indicate citations from the article corresponding to the start node to the article corresponding to the arrival node.}
\label{fig:network example}
\end{figure}

Based on the citation information provided by the reference list, we can construct a citation network. In particular, we only consider the citation relationships among the articles in our dataset. In the citation network, nodes represent articles and directed edges indicate the citation relationship between two articles. If article $A$ cites article $B$, then there is a directed edge pointing from $A$ to $B$. Since earlier published articles cannot cite itself and later published articles, the citation network is a directed acyclic graph (DAG) and there is no self-loop. To better illustrate the structure and elements of the citation network, we present a sub-graph. We begin by searching the dataset using the keyword ``adaptive lasso'', which yields a total of 50 articles. Next, we obtain the network of the citation relationships based on these 50 articles and plot the connected parts (31 nodes), as shown in Figure \ref{fig:network example}. The bigger the size of the node, the more citations it has from the 31 articles. It can be seen that there are two articles that have been cited more frequently by other articles, namely ``The adaptive Lasso and its oracle properties" and ``Adaptive Lasso for sparse high-dimensional regression models". 

Subsequently, we calculate the density of our citation network. The theoretical density of a DAG is $2m/\{n(n-1)\}$, where $m$ represents the number of edges, and $n$ represents the number of nodes. Our network comprises 190,381 nodes and 1,087,277 directed edges, leading to a network density of 0.0060\%. The densities of several citation networks constructed by other scholars or academic institutions are listed below. \cite{shen2014modeling} construct a citation network containing all articles and citations published by the American Physical Society between 1893 and 2009. This network consists of 463,348 nodes and 4,710,547 directed edges, resulting in a network density of 0.0044\%. The INSPIRE, as reported above, has a network density of 0.0041\%. Both of the two networks mentioned above are networks in the field of physics. The densities of networks of similar scale in a specific discipline reflect the validity of our network construction in the field of statistics.

\section{Recommendation System}\label{sec4}

In this section, we introduce our recommendation system, which is organized into three main components. Specifically, Section \ref{Candidate article Set Selection} describes the selection of a candidate set of articles from which recommended articles are generated. Section \ref{Similarity Calculation} describes our approach to computing similarity, which incorporates both content-based and network-based measurements. Section \ref{Recommendation Mechanism} provides a detailed overview of the operation process of our recommendation system, starting from the input article and ultimately presenting the recommended articles. 

\subsection{Candidate article Set Selection}
\label{Candidate article Set Selection}

As mentioned above, it is computationally expensive to compare every article with the input article for each recommendation. To find suitable candidate articles for a specific article, we decide to look for its neighbors in the citation network as well as the neighbors of those neighbors. In directed networks, there are two types of neighbors, that is, in-neighbors and out-neighbors \citep{maserrat2010neighbor}. If article $A$ cites article 
$B$, then article $A$ is the in-neighbor of article $B$, and article $B$ is the out-neighbor of article $A$. An out-neighbor query on article $A$ is to find the set of out-neighbors of article $A$. Similarly, an in-neighbor query on article $B$ is to search for all in-neighbors of article $B$. Given a specific article, the articles that are most directly related to it are its in-neighbors and out-neighbors. 

\begin{figure}[!htbp]
\centering
\includegraphics[scale=0.35]{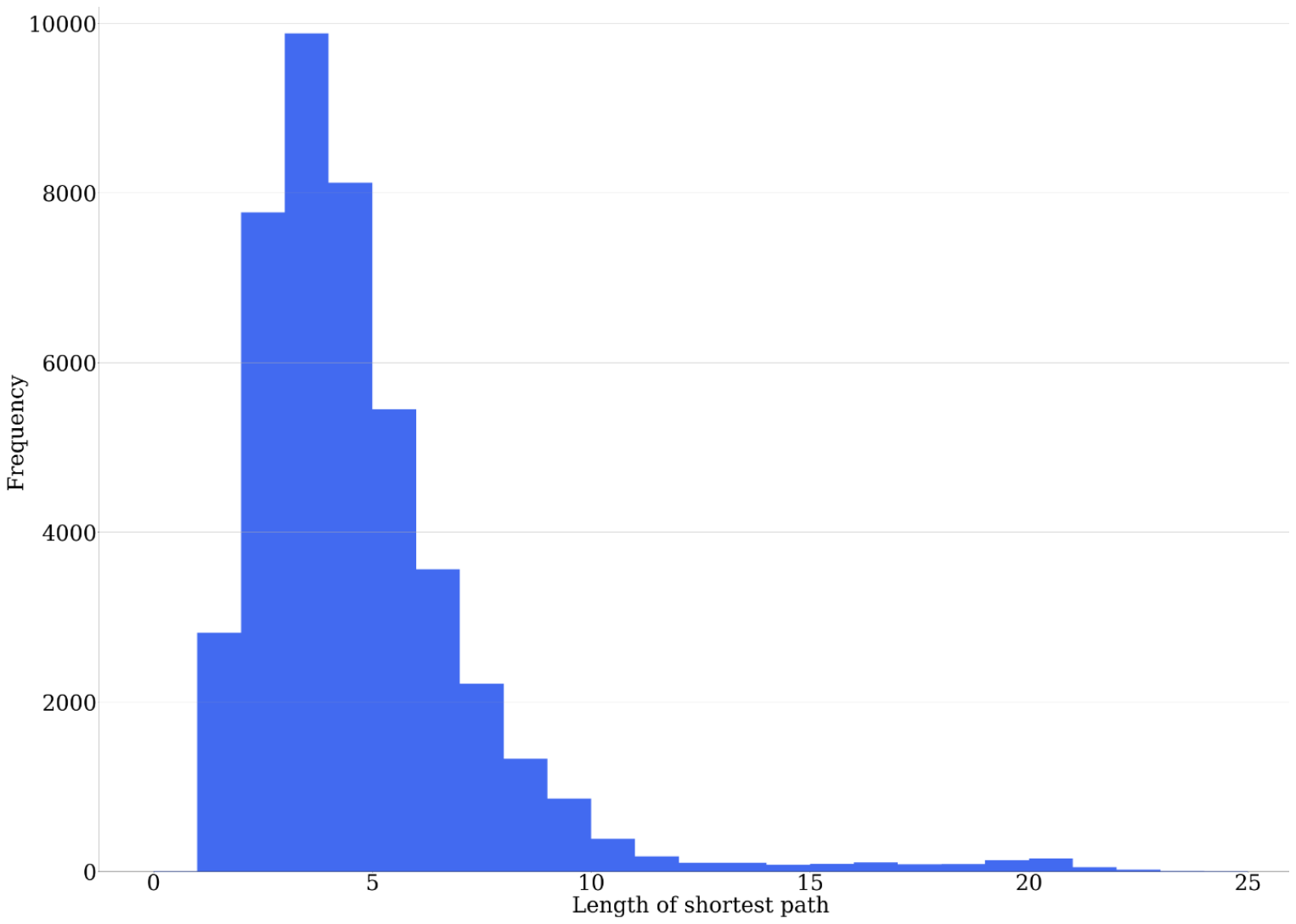}
\caption{Histogram of the lengths of the existing shortest paths from all other nodes to article\_Lasso. Almost half of the nodes were able to reach this highly cited node within 3 steps.}
\label{fig:diji}
\end{figure}

In-neighbors include recent articles that are most relevant to this article. Since our dataset is relatively new, a single in-neighbor query is sufficient to find cutting-edge articles. However, conducting only one out-neighbor query for a particular article may not be sufficient to capture classical and relevant articles. For example, the node with the largest in-degree in the citation network is the article titled ``Regression Shrinkage and Selection via the Lasso", published in 1996. For convenience, we refer to this article as article\_Lasso in the following text. 
We compute the lengths of the shortest paths from all other nodes to article\_Lasso according to the algorithm proposed in \cite{Dijkstra1959ANO}. In a citation network, the length of the shortest path from article $A$ to article $B$ equals the minimum number of steps required to cite from $A$ to $B$. For example, if $A$ directly cites $B$, that is, $A$ is the in-neighbor of $B$, the length of the shortest path from $A$ to $B$ is 1. If $A$ does not directly cite $B$, and if it is possible to find that $A$ cites $C$ and $C$ cites $B$, then the length of the shortest path from $A$ to $B$ is 2, and so on. If there is no path from $A$ to $B$, obviously there does not exist a shortest path. The distribution of the lengths of the existing shortest paths is shown in Figure \ref{fig:diji}. As can be observed, the nodes that reach article\_Lasso at the third step are the most numerous. In addition, we find that within the articles that have the shortest paths to article\_Lasso, almost 50\% of the articles can reach article\_Lasso in three steps.

\begin{figure}[!htbp]
\centering
\includegraphics[scale=0.5]{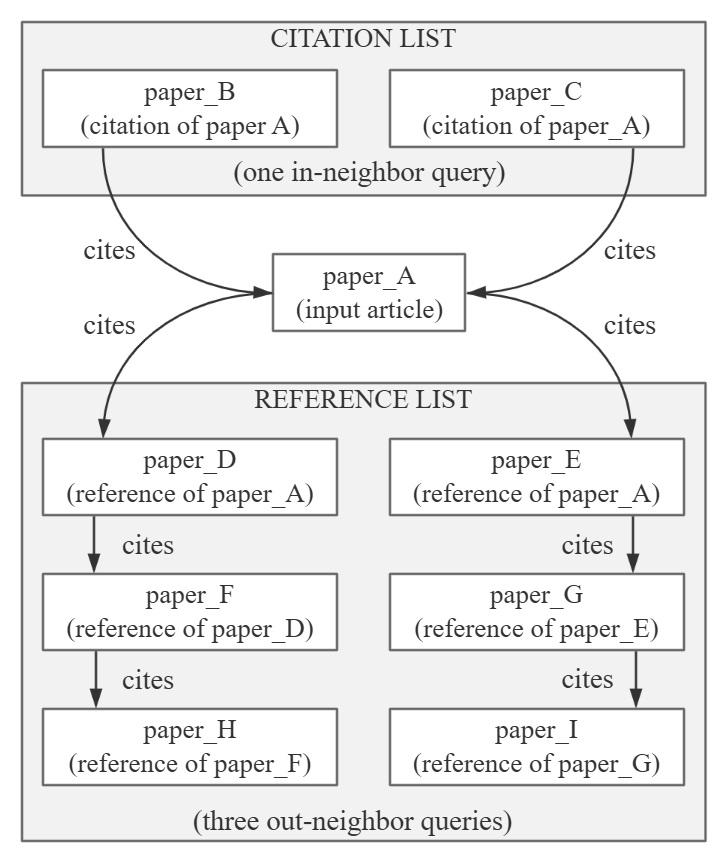}
\caption{Schematic diagram of one in-neighbor query and three out-neighbor queries. The in-neighbor query provides a CITATION LIST, and the out-neighbor queries offer a REFERENCE LIST.}
\label{fig:candidate}
\end{figure}

Inspired by the above observations, we refer to the set of articles within three out-neighbor queries as the REFERENCE LIST, and the set of articles within one in-neighbor query as the CITATION LIST. The REFERENCE LIST and the CITATION LIST comprise our set of candidate articles, as shown in Figure \ref{fig:candidate}. In subsequent article recommendations, we only calculate the similarity between candidate articles and the input article. A detailed description of the similarity measures can be found in the following subsection.  

\subsection{Similarity Calculation}
\label{Similarity Calculation}

Given two articles, we construct content-based and network-based similarity measurements between them. The content-based similarity measurements are determined by comparing the similarities of the abstracts and titles of the two articles. We first utilize the {\it text-embedding-3-small model} interface provided by OpenAI to obtain their abstract vectors. Text embedding is a process that converts text into numerical vectors, which represent the semantic meaning of the text in a way that statistical models can understand and process \citep{bengio2003neural}. The {\it text-embedding-3-small model} uses the transformer architecture, a deep learning model commonly used to process sequential data. In transformer, the self-attention mechanism allows the model to process each word while considering all words in the entire input sequence. This mechanism allows the model to better understand context and semantics. By default, the {\it text-embedding-3-small model} returns an embedding of 1,536 dimensions, which is sufficient for summarized text vectorization of around 200 words. Thus, each abstract is converted into a vector of 1,536 dimensions. We then compute the cosine similarity between the two articles. Denote $v(A) \in \mathbb{R}^{1536}$ as the embedding vector of article $A$.
Cosine similarity is a measure of how similar two vectors are in terms of direction, which is widely used to compute similarity between texts \citep{huang2014refseer,bhagavatula2018content,sterling2021combining}. Specifically, for article $A$ and article $B$, their cosine similarity is defined as
$$\text{cos}\big(v(A),v(B)\big) = \frac{v(A)^\top v(B)}{\|v(A)\| \|v(B)\|},$$
where $\|\cdot\|$ is the $L_2$ norm of a vector.

A small number of articles do not have abstracts, and we set the similarity of abstracts to the mean of the other articles if one of the two articles does not have an abstract. As a complementary measurement, we compute the similarity of titles. Differences in content characteristics and semantic expressions between abstracts and titles determine that they are suitable for different vectorization methods. Abstracts are usually semantically rich, contextually relevant and long, so they are more suitable for word embedding vectorization (e.g., Word2Vec or BERT), which can capture the semantic relationships between words and contextual information to better reflect the complex semantics of abstracts \citep{pittaras2021text}. In contrast, titles are usually shorter, keyword-based, and have relatively simple semantics, making them more suitable for word frequency vectorization (e.g., TF-IDF or bag-of-words (BoW) models) \citep{selva2021review}. Word frequency vectorization is simple and efficient to compute, can quickly capture the importance of keywords, and is suitable for handling similarity computation for short texts \citep{ajose2020performance}. We summarize the characteristics of abstracts and titles in Table \ref{table:Characteristics}. Consequently, we select the BoW model to vectorize titles. For two titles, the BoW model will segment them into two-word lists, and then all occurrences of words are counted to construct a vocabulary. Each word in the vocabulary corresponds to a feature (i.e., a dimension of the vector). For each title, the BoW model counts the number of occurrences of each word in the vocabulary in its title and uses these counts as feature values. Eventually, each title is represented as a sparse vector whose length is equal to the size of the vocabulary list, and each element represents the word frequency of the corresponding word.

%\textcolor{violet}{For example,} 
Another important similarity measurement is constructed using our citation network. For each article in the two lists, we calculate the number of common neighbors between that article and the input article. Specifically, if we have two articles A and B, their common neighbors are the articles that have directed edges coming from both A and B. This measurement is often used in various graph-based algorithms to determine the similarity or connection strength between nodes. By applying max-min regularization to the number of node co-connections, we then derive the node similarity of the citation network.

\begin{table}[htp]
\caption{\textbf{Characteristics of abstracts and titles, including differences in semantic complexity, text length, semantic relevance, and popular vectorisation methods.}} % 表格标题
\label{table:Characteristics}
\centering % 表格居中
\begin{tabular}{lcc} % 表格有三列
\toprule % 表格顶部横线
Characteristics & Abstract & Title \\
\midrule % 表格中间横线
Semantic complexity & high & low \\
Text length & long  & short \\
Semantic relevance & Strong contextual relevance & Strong keyword relevance \\
Popular vectorisation methods & \makecell{word embedding vectorisation \\(Word2Vec, BERT, etc.)} & \makecell{word frequency vectorisation \\ (TF-IDF, BoW models)} \\
\bottomrule % 表格底部横线
\end{tabular}
\end{table}

\subsection{Recommendation Mechanism}
\label{Recommendation Mechanism}

\begin{figure}[!htbp]
\centering
\includegraphics[width=\textwidth,height=0.4\textheight]{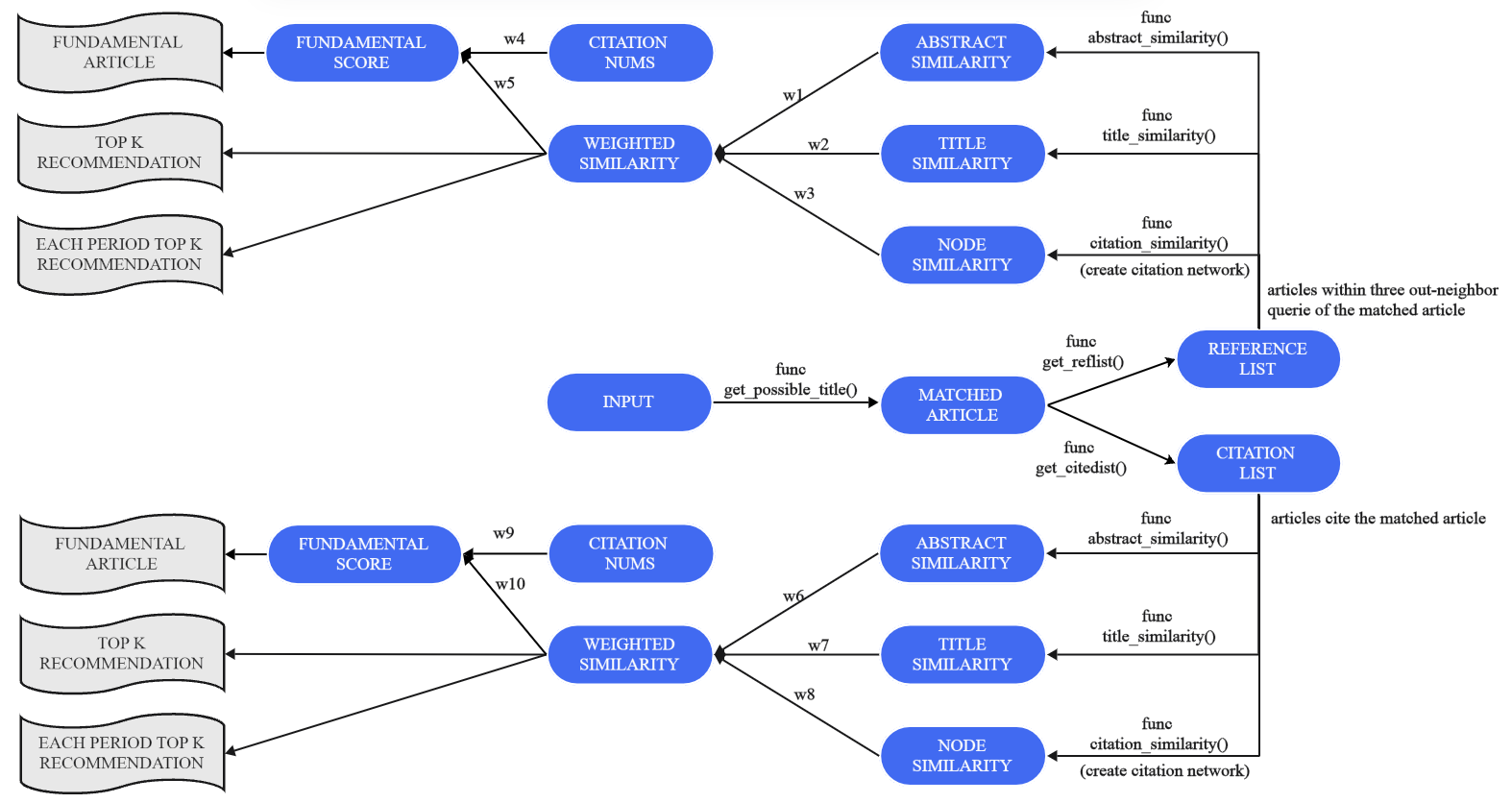}
\caption{Flowchart of our recommendation mechanism, containing the functions involved (above the straight line of arrows) and important objects such as the computed similarity metrics (ellipses) and the final output (flags). }
\label{fig:recommendation mechanism}
\end{figure}

The flowchart of our recommendation method is shown in Figure \ref{fig:recommendation mechanism}, which is divided into three main parts, i.e., top, middle and bottom. Among them, the middle part centers on obtaining the candidate article set, i.e., the REFERENCE LIST and the CITATION LIST. For each title entered, our method first searches the full dataset to see if there is a matched article. This search supports fuzzy matching, which tolerates spelling errors to a certain extent. Furthermore, our recommendation mechanism also supports keyword matching recommendation. If the user does not have an explicit article that he/she wants to query, then he/she can enter a keyword or a key phrase. The recommendation method will match relevant articles for the input keyword, thus achieving a more universal article recommendation function. The user can select the most relevant article from the matched search results for further recommendation, i.e., the matched article. After obtaining the matched article in our dataset, the candidate article set from which recommended articles are generated. As previously stated, our article recommendation system limits the candidate article set to the REFERENCE LIST and CITATION LIST. The top part calculates similarities of articles in the REFERENCE LIST to the matched article and makes similar article recommendations based on the weighting of the three similarities. Similar articles are also weighted with the number of citations needed to obtain a highly cited similar article, i.e., a fundamental article. The CITATION LIST is passed to the bottom section, which also performs similarity calculations and weighting to obtain similar article recommendations and highly cited similar articles. The process of ranking and recommending articles in the REFERENCE LIST and the CITATION LIST is carried out separately. However, except for the difference in the scope of the recommended articles, the ranking and recommendation process is exactly the same. Therefore, in the following detailed description of the mechanism, we will only expand the process of sorting and recommending articles in the REFERENCE LIST.

In the top or bottom part, the literature recommendation mechanism consists of four steps which are detailed in the following. 
\begin{enumerate}[Step 1:]
    \item[Step 1:] Calculate the abstract similarity, title similarity, and node similarity between each article in the candidate article set and the matched article according to the methods introduced in Section \ref{Similarity Calculation}.

    \item[Step 2:] Users can assign weights to different similarities based on their preferences. The weighted similarity can be calculated as follows:
    $$
    \text{weighted-sim} = w_1 \times \text{abstract-sim} + w_2 \times \text{title-sim} + w_3 \times \text{node-sim},
    $$
    \vspace{-0.7cm}
    $$
    \text{s.t.,}\ \sum_{i=1}^3 w_i =1,\ w_i \geq 0,\ \text{for}\ i=1,2,3,
    $$
    where $\text{weighted-sim}, \text{abstract-sim}, \text{title-sim},$ and $\text{node-sim}$ denote the weighted similarity, abstract similarity, title similarity, and node similarity, respectively. For example, to prioritize articles with higher title similarity, one can increase the weight $w_2$ while correspondingly reducing other weights to satisfy the summation constraint.

    \item[Step 3:] Recommend articles based on weighted similarity in two approaches. One is to recommend the top $k$ articles with the highest weighted similarity overall. The other focuses on recommending the top $k$ articles with the highest weighted similarity for each 5-year period, recognizing that there may be some differences in articles on the same topic between different periods.

    \item[Step 4:] Further consider the number of citations of articles in the set of candidate articles. articles with a high number of citations are generally considered to be well-recognized articles within the field. Consequently, normalize the number of citations through max-min regularization. Then, take the normalized number of citations and perform a weighted average with the weighted similarity to obtain the recommendation score, which can be calculated as follows:
    $$
    \text{fundamental score} = w_4 \times \text{normalized num of citations} + w_5 \times \text{weighted-sim},
    $$
    \vspace{-0.7cm}
    $$
    \text{s.t.,}\ w_4+w_5=1,\ w_4,\ w_5 \geq 0.
    $$
    These weights also allow user-defined assignments. When recommending articles from CITATION LIST, one can use weights $w_6$ to $w_{10}$ which are defined in Figure \ref{fig:recommendation mechanism} instead. The top $k$ articles with the highest fundamental score are considered the fundamental article. 
\end{enumerate}

\section{Evaluation}\label{sec5}

In this section, we evaluate our article recommendation method in two ways. In Section \ref{5.1}, we select 10 review articles and use our method to reconstruct the reference lists of the review articles. The hit rate is used as an evaluation metric to assess the usability of our method. In Section \ref{5.2}, we use recall metrics including hit@1, hit@5, hit@10, recall@20 and the best rank. Similarly, we evaluate the ability of our method to reconstruct references of non-review articles by reconstructing their reference lists.
 
\subsection{Review Article Reference Reconstruction} \label{5.1}

The objective of this section is to examine the efficiency of the literature recommendation method, particularly focusing on its usefulness in helping authors writing review articles. To this end, we adopt the existing approach of reconstructing the reference lists of review articles. However, our dataset encompasses data from only 70 journals, which means some references may be absent from our dataset. Consequently, for each review article, we determine the number of references available in our database, denoted as $k$. We then proceed to recommend $k$ similar articles. The hit rate is calculated as the proportion of these $k$ recommended articles that are indeed present in the reference list of the review article. The evaluation results for the 10 selected review articles are shown in Table \ref{tab:review_articles}. The average hit rate reaches 0.7, which is comparable to the hit rate of existing work \citep{sterling2021combining}. 

Taking the first article in Table \ref{tab:review_articles} as an example, we observe the characteristics of references not hit by the recommendations generated by our method. The article ``Methodologies in spectral analysis of large dimensional random matrices: A review'' provides an extensive examination of techniques used in analyzing the eigenvalues and eigenvectors of large random matrices. It covers foundational aspects of random matrix theory, various analytical and computational methodologies, and applications across physics, finance, and data science. The review underscores the importance of spectral analysis in modeling complex systems and highlights future research directions in handling high-dimensional data and improving computational algorithms. Two references of this article existing in our dataset are not recommended, neither of which has an abstract. We have set the abstract similarity between articles without abstracts and the input article to the mean abstract similarity of articles with abstracts and the input article. However, the abstract similarities of the articles without abstracts are still affected.

\begin{table}[ht]
\centering
\caption{10 selected review articles and their hit rates. The term ``\# of Refs in Dataset" corresponds to the number of references in each review article that are included in our dataset, while ``\# of total references" refers to the total number of references in each review article.}
\label{tab:review_articles}
\begin{tabular}{@{}l c c  c c@{}}
\toprule
\textbf{Title} & \textbf{Year} & \textbf{\makecell{\# of \\ Refs in \\Dataset}} & \textbf{\makecell{\# of \\Total\\ Refs}} & \textbf{\makecell{Hit\\Rate \\(\%)}} \\
\midrule
\makecell[l]{Methodologies in spectral analysis of large \\dimensional random matrices, a review} & 1988 & 15 & 54 & 86.67 \\
\makecell[l]{Analyzing unreplicated factorial experiments:\\ A review with some new proposals} & 1998 & 12 & 45 & 83.33 \\
\makecell[l]{A review of the development and application\\ of recursive residuals in linear models} & 1996 & 13 & 62 & 76.92 \\
\makecell[l]{From simple structure to sparse components: A review} & 2014 & 23 & 56 & 73.91 \\
\makecell[l]{A review and a synthesis of the fast fourier transform\\ algorithms for exact analysis of discrete data} & 1997 & 18 & 54 & 72.22 \\
Counting process models for life history data: a review & 1985 & 38 & 154 & 65.79 \\
\makecell[l]{The calculus for factorial arrangements: \\A review and bibliography} & 1989 & 20 & 126 & 65.00 \\
\makecell[l]{Models for paired comparison data: \\A review with emphasis on dependent data}& 2012 & 35 & 120 & 57.14 \\
A review on anisotropy analysis of spatial point patterns & 2018 & 20 & 50 & 55.00 \\
Methods for national-population forecasts - a review & 1986 & 11 & 101 & 54.55 \\

\bottomrule
\end{tabular}
\end{table}

\subsection{Evaluation on Non-Review Articles}\label{5.2}

Unlike the reconstruction of reference lists for review articles in Section \ref{5.1}, the evaluation in this section is conducted on non-review articles. The writing of non-review articles differs significantly from that of review articles. Non-review articles typically have fewer references. Moreover, existing research indicates that articles published in a particular journal tend to cite historical articles from that journal, which may result in a higher number of reference articles included in our dataset. Therefore, we select four evaluation metrics, including hit@1, hit@5, and hit@10, to assess whether our system can recommend articles required by scholars with high priority. The metric ``hit@1" corresponds to the hit rate of recommended articles when recommending 1 article, and so on. Additionally, recall@20 is used in cases of extensive recommendations to evaluate whether our method can recommend as many required articles as possible. The metric ``recall@20" corresponds to the proportion of recommended articles existing in the reference list of the non-review article when recommending 20 articles.

The assessment is conducted on 1,500 randomly selected non-review articles. The average hit rate for the first-ranked article (hit@1) is 0.85, which means that 85\% of the first recommended articles for non-reviewed articles are included in their reference list. The hit@5 is 0.44 and hit@10 is 0.28, indicating that our method is capable of accurately recommending articles required by researchers. By looking at the performance of the evaluation scores of different articles, we find that some of the articles have only a few references in the dataset, for example, when an article has only two references. Thus, hit@5 reaches at most 0.4, which results in a decrease in the mean value of the indicator. Based on the results of the existing works, we achieve a higher hit rate when recommending 1 to 5 articles compared to \cite{yang2013task}, and achieve a comparable hit@10 compared to \cite{pornprasit2022enhancing}. The average recall rate when recommending 20 articles (recall@20) is 0.76 compared to 0.5 \citep{yang2013task}. This means that when 20 articles are recommended, 76\% of the real references can be recommended. In addition, both hit rates and recall rates have declined in the last few years of recommendations. Increasingly long titles and an increasing number of publications make it more difficult to improve recommendation coverage.

\begin{table}[ht]
\centering
\caption{Ten selected non-review articles with their hit rates and recall rate. The term ``hit@1" corresponds to the hit rate of recommended articles when recommending 1 article, while ``recall@20" corresponds to the proportion of recommended articles existing in the reference list of the non-review article when recommending 20 articles.}
\label{tab:non_review_articles}
\begin{tabular}{@{}l c c c c c c c@{}}
\toprule
\textbf{Title} & \textbf{Year} & \textbf{\makecell{\# of \\ Refs in \\Dataset}} & \textbf{\makecell{\# of \\Total\\ Refs}} & \textbf{\makecell{hit\\@1}} & \textbf{\makecell{hit\\@5}} & \textbf{\makecell{hit\\@10}} & \textbf{\makecell{recall\\@20}} \\
\midrule

\makecell[l]{Minimax properties of $M-, R-$ and $L$-\\estimators of location in levy neigh-\\bourhoods} & 1989 & 3 & 15 & 1.00 & 0.60 & 0.30 & 1.00\\
\makecell[l]{A note on unconditional properties of a\\ parametrically guided Nadaraya-Watson\\ estimator} & 1998 & 4 & 8 & 1.00 & 0.60 & 0.40 & 1.00\\

\makecell[l]{Asymptotics for trimmed k-means and \\associated tolerance zones} & 1999& 8 & 26 & 0.00 & 0.80 &0.70 &1.00 \\

\makecell[l]{A martingale approach to the copula-graphic\\ estimator for the survival function under\\ dependent censoring} & 2001 & 11 & 26 & 1.00 & 0.80 & 0.60 & 0.82\\

\makecell[l]{Fisher information in ordered randomly \\censored data with applications to char-\\acterization problems} & 2003& 12& 19 & 1.00 & 0.80 &0.70 &0.58 \\

\makecell[l]{Orthogonal-column Latin hypercube designs\\ with small samples} & 2009 & 13 & 25 & 1.00 & 0.80 &0.90 &0.85 \\
\makecell[l]{Sparse model selection under heterogeneous\\ noise: Exact penalisation and data-driven\\ thresholding} & 2014 & 8 & 25& 1.00 & 1.00 & 0.70 & 0.88\\

\makecell[l]{Parameter estimation for the logistic reg-\\ression model under case-control study} & 2016 & 7 & 12 & 1.00 & 0.80 &0.60 &0.86 \\

\makecell[l]{On the estimation of nested Archimedean\\ copulas: A theoretical and an experimental\\ comparison} & 2018 & 10 & 28 & 1.00 & 0.60 &	0.60 &0.80 \\

\makecell[l]{Ultrahigh-dimensional generalized additive\\ model: Unified theory and methods} & 2021 & 30 & 42 & 0.00 & 0.60 & 0.40 & 0.17\\

\bottomrule
\end{tabular}
\end{table}

\section{Conclusions and Future Directions}\label{sec6}

Literature recommendation plays a crucial role in academic research, especially in the current era of information overload. Traditional methods of literature search, such as keyword-based search engines, can help to some extent, but often return a large number of irrelevant results, consuming significant time and effort. The ability to effectively discover the most relevant and influential articles has become a critical issue.
In this work, we construct a large citation network which includes 70 journals across statistics, econometrics, and computer science. Unlike purely network-based or content-based approaches, our hybrid method effectively captures both structural relationships and semantic similarities among academic articles. In addition, the inclusion of abstracts embedded through a large language model substantially boosts recommendation quality, enabling more accurate identification of relevant literature. When new articles are added, our embedding approach only requires embedding the abstracts of the new articles, greatly reducing computational costs and avoiding the need to reprocess existing data. The experimental results demonstrate the effectiveness of our approach in identifying influential articles.

In the future, several directions can be explored to enhance the literature recommendation. First, incorporating additional types of metadata, such as co-authorship networks, journal impact factors, and trending research topics, could help capture more comprehensive relationships between articles. Second, while we currently use the number of common neighbors to measure similarity in citation patterns, there is potential to integrate more advanced techniques like deep learning and graph neural networks to better model complex relationships within citation networks.  Finally, as new articles are published, it will be essential to develop a system that can automatically add them to the dataset and calculate their similarity to existing articles. Additionally, by considering the research history, preferences, and behavior of users, the system could provide more adaptive and personalized recommendations.

\section*{Acknowledgments}

The research of Rui Pan is supported by the National Natural Science Foundation of China (No. 72471254), the Program for Innovation Research, the Disciplinary Funds and the Emerging Interdisciplinary Project of Central University of Finance and Economics. The research of Hansheng Wang is supported by National Natural Science Foundation of China (Nos. 12271012, 72495123).

\newpage

\begin{appendices}

\section{Appendix A}\label{secA1}

\begin{table}[!htbp]
        \tiny%字号
	\caption{The 70 journals we select, whith can be mainly divided into three categories, i.e., statistics, econometrics, and computer science. The journals are listed in alphabetical order, accompanied by the time range of the articles crawled and the number of articles included.}
        \vspace{0.5cm}
	\label{Tab:journals}
        \newcommand{\tabincell}[2]{\begin{tabular}{@{}#1@{}}#2\end{tabular}}
        %\begin{center}
        \begin{tabular}{cccc}
        \hline
        \hline
        \textbf{ID} & \textbf{Journal} & \textbf{Period} & \textbf{\# of articles}\\
        \hline
	1 & ACM Transactions on Knowledge Discovery from Data  & 2009-2022 & 708\\
        2 & Advances in Data Analysis and Classification  & 2007-2022 & 391\\
        3 & American Statistician  & 1981-2022 & 1,957 \\
        4 & Annals of Applied Statistics & 2007-2022 & 1,319 \\
        5 & Annals of Statistics  & 1981-2022 & 4,082\\
        6 & Annals of the Institute of Statistical Mathematics & 1981-2022 & 1,976\\
        7 & Annual Review of Statistics and Its Application  & 2014-2022 & 127\\
        8 & Artificial Intelligence  & 1981-2022 & 2,575\\
        9 & Australian \& New Zealand Journal of Statistics  & 1998-2022 & 769 \\
        10 & Bernoulli & 1997-2022 & 1,638\\
        11 & Bioinformatics & 1997-2022 & 15,577 \\
        12 & Biometrics & 1981-2022 & 5,043\\
        13 & Biometrika & 1981-2022 & 3,418\\
        14 & Biostatistics & 2002-2022 & 1,161\\
        15 & Communications in Statistics-Simulation and Computation & 1981-2022 & 5,345\\
        16 & Communications in Statistics-Theory and Methods & 1981-2022 & 11,151\\
        17 & Computational Statistics & 1994-2022 & 1,638\\
        18 & Computational Statistics \& Data Analysis & 1983-2022 & 5,902\\
        19 & Data Mining and Knowledge Discovery  & 1997-2022 & 957\\
        20 & Econometrica & 1981-2022 & 1,994\\
        21 & Electronic Journal of Statistics & 2007-2022 & 1,613\\
        22 & IEEE Transactions on Cybernetics & 2013-2022 & 4,628\\
        %23 & IEEE Transactions on Image Processing & 1992-2022 & 9,137\\
        23 & IEEE Transactions on Information Theory & 1981-2022 & 11,913\\
        24 & IEEE Transactions on Knowledge and Data Engineering & 1992-2022 & 4,356\\
        25 & IEEE Transactions on Neural Networks & 1990-2012 & 2,754\\
        26 & IEEE Transactions on Pattern Analysis and Machine Intelligence & 1981-2022 & 4,176\\
        %28 & IEEE-ACM Transactions on Computational Biology and Bioinformatics & 2004-2022 & 2,611\\
        27 & International Statistical Review & 1981-2022 & 880\\
        28 & Journal of Applied Statistics & 1993-2022 & 3,532\\
        29 & Journal of Business \& Economic Statistics & 1985-2022 & 1,818\\
        30 & Journal of Computational and Graphical Statistics & 1997-2022 & 1,573\\
        31 & Journal of Econometrics & 2001-2022 & 1,518\\
        32 & Journal of Machine Learning Research & 2001-2022 & 995\\
        33 & Journal of Multivariate Analysis & 1981-2022 & 4,229\\
        34 & Journal of Nonparametric Statistics & 1998-2022 & 1,105\\
        35 & Journal of Statistical Computation and Simulation & 1981-2022 & 3,659\\
        36 & Journal of Statistical Planning and Inference & 1981-2022 & 6,616\\
        37 & Journal of Statistical Software & 2005-2022 & 1,170\\
        38 & Journal of Survey Statistics and Methodology & 2018-2022 & 230\\
        39 & Journal of the American Statistical Association & 1981-2022 & 5,536\\
        40 & Journal of the Royal Statistical Society Series A-Statistics in Society & 1981-2022 & 1,443\\
        41 & Journal of the Royal Statistical Society Series B-Statistical Methodology & 1981-2022 & 1,202\\
        42 & Journal of the Royal Statistical Society Series C-Applied Statistics & 1981-2022 & 1,585\\
        43 & Journal of the Royal Statistical Society Series D-The Statistician & 1981-2022 & 551\\
        44 & Journal of Time Series Analysis & 2000-2022 & 996\\
        45 & Machine Learning & 1990-2022 & 1,822\\
        46 & Neural Computation & 1989-2022 & 3,073\\
        47 & Neural Computing and Applications & 1995-2022 & 7,717\\
        48 & Neural Networks & 1988-2022 & 4,712\\
        49 & Neurocomputing & 1992-2022 & 17,238\\
        50 & Physica A: Statistical Mechanics and its Applications & 1981-2022 & 12,140\\
        51 & R Journal & 2009-2022 & 582\\
        52 & Review of Economics and Statistics & 1981-2022 & 2,757\\
        53 & Scandinavian Journal of Statistics & 1981-2022 & 1,680\\
        54 & Spatial Statistics & 2012-2022 & 568\\
        55 & Stat & 2016-2022 & 333\\
        56 & Stata Journal & 2001-2022 & 862\\
        57 & Statistica Sinica & 1991-2022 & 2,333\\
        58 & Statistical Analysis and Data Mining & 2013-2022 & 337\\
        59 & Statistical Methodology & 2011-2017 & 273\\
        60 & Statistical Methods and Applications & 2007-2022 & 538\\
        61 & Statistical Methods in Medical Research & 1992-2022 & 1,751\\
        62 & Statistical Modelling & 2001-2022 & 512\\
        63 & Statistical Papers & 1993-2022 & 1,648\\
        64 & Statistical Science & 1993-2022 & 732\\
        65 & Statistics & 1993-2022 & 1,433\\
        66 & Statistics \& Probability Letters & 1984-2022 & 8,849\\
        67 & Statistics and Computing & 1993-2022 & 1,541\\
        68 & Statistics in Medicine & 1982-2022 & 8,956\\
        69 & Technometrics  & 1981-2022 & 1,481\\
        70 & Test & 1997-2022 & 748\\
        \hline
	\end{tabular}
%\end{center}
\end{table}

\end{appendices}

\newpage

\bibliography{sn-bibliography}% common bib file

\end{document}